\documentclass[nofootinbib,twocolumn,aps,amsmath,amssymb]{revtex4}
\usepackage{graphics,bm}
\usepackage{epsfig}
\usepackage{graphicx}

\def\sumdiff{\hbox{$\Delta$}\!\!\!\!\!\!\int}

\newcommand{\beq}{\begin{equation}}
\newcommand{\eeq}{\end{equation}}
\newcommand{\bqa}{\begin{eqnarray}}
\newcommand{\eqa}{\end{eqnarray}}
\newcommand{\simorder}{\raisebox{-4pt}{$\, \stackrel{\textstyle >}{\sim} \,$}}
\newcommand{\simordertwo}{\raisebox{-4pt}{$\, \stackrel{\textstyle<}{\sim}\,$}}
\begin{document}

\parindent=20pt
\parskip=10pt
\pagestyle{plain}
\font\tenrm=cmr10
\def\sumint{\hbox{$\sum$}\!\!\!\!\!\!\int}
\def\square{\vcenter{\vbox{\hrule height.4pt
          \hbox{\vrule width.4pt height4pt
          \kern4pt\vrule width.3pt}\hrule height.4pt}}}
\def\boxx{\square}
\def\isumint{\hbox{${\scriptstyle \Sigma}$}\!\!\!\!\!\int}
\def\isumdiff{\hbox{${\scriptstyle \Delta}$}\!\!\!\!\!\int}
\def\ranglec{\rangle_{\!\!c}}
\def\ranglex{\rangle_{\!\!x}}
\def\ranglecx{\rangle_{\!\!c,x}}

\title{Thermodynamics of $O(N)$ sigma models: $1/N$ corrections}
\author{Jens O. Andersen} \affiliation{Nordita,\\
Blegdamsvej 17, DK-2100 Copenhagen, Denmark}
\email{jensoa@nordita.dk}
\author{Dani\"el Boer and Harmen J. Warringa} \affiliation{Department
of Physics and Astronomy, Vrije Universiteit, De Boelelaan 1081, 1081
HV Amsterdam, The Netherlands}
\email{dboer@nat.vu.nl; harmen@nat.vu.nl}
\date{\today}

\begin{abstract}
The thermodynamics of the $O(N)$ linear and nonlinear sigma models in
3+1 dimensions is studied. We calculate the pressure to
next-to-leading order in the $1/N$ expansion and show that at this
order, temperature-independent renormalization is only possible 
at the minimum of the effective potential.
The $1/N$ expansion is found to be a good expansion for $N$ as low as 4, 
which is the case relevant for low-energy QCD phenomenology. We consider the
cases with and without explicit symmetry breaking. 
We show that previous next-to-leading order calculations of the 
pressure are either breaking down in the temperatures of interest, or 
based on unjustifiable high-energy approximations.
\end{abstract}

\maketitle

\section{Introduction}
It is well-known that although 
the QCD Lagrangian possesses a chiral symmetry in the limit 
of zero quark masses, the true QCD ground state does not respect 
this symmetry. The chiral symmetry is spontaneously broken by 
quantum effects.
To be specific, the QCD Lagrangian with $N_f$ massless quarks has a global  
$SU(N_f)_L\times SU(N_f)_R$ symmetry, which for the ground state at low
temperatures is broken down to an $SU(N_f)_V$ symmetry. According to 
Goldstone's theorem, there is a massless, spinless particle for each 
generator of a broken global continuous symmetry. In this case this implies 
the occurrence of $N_f^2-1$ Goldstone bosons. In phenomenological
applications $N_f$ is either two or three, 
and one also has to take into account the
explicit symmetry breaking due to the nonzero quark masses. Both the
spontaneous and the explicit chiral symmetry breaking are 
apparent in the low-energy hadronic 
particle spectrum, where the expected number of 
relatively light mesons is observed (e.g.\ the three pions for $N_f=2$). At 
sufficiently 
high temperatures one expects the chiral symmetry to be restored and 
lattice simulations of QCD suggest that this happens 
at a temperature of approximately 150 MeV depending on the number of quarks
and their masses.
Heavy-ion collisions at RHIC and LHC are expected to reach such temperatures 
and will allow experimental studies of the deconfined, chirally symmetric 
phase of QCD~\cite{heavyion3}. 

Apart from using lattice simulations it has not yet been possible to calculate
thermodynamic properties, such as the pressure, from QCD in the
low-temperature hadronic phase. However, this can
be done using low-energy effective theories. Such effective 
theories for the low-energy particle spectrum, involving both mesons and
baryons and displaying the above-mentioned pattern of chiral
symmetry breaking, were constructed before QCD itself.
In the case of two flavors, the situation is the simplest, since one can 
exploit the fact that the $SU(2)_L\times SU(2)_R$ symmetry 
is locally isomorphic to $SO(4)$.
If baryons (nucleons in this case) are not included
the simpler $O(4)$ linear sigma model can be used as a low-energy 
effective theory for describing the dynamics of three pion fields and one 
sigma field. These four fields form a four-dimensional vector $\phi$ 
in the fundamental representation of $O(4)$.
At low temperature, the $O(4)$ symmetry is 
spontaneously broken down to $O(3)$, where the sigma field 
acquires a vacuum expectation
value and the three pions are interpreted as the Goldstone bosons. 
For $N_f >2$ there is no connection between the $SU(N_f)_L\times SU(N_f)_R$
model and the $O(N)$ linear sigma model, but the latter has been
studied in great detail for general $N$ due to its relevance
to spin models.

In this paper the $O(N)$ linear sigma model (LSM) and  
$O(N)$ nonlinear sigma model (NLSM) in 3+1 dimensions will be studied at
finite temperature and to next-to-leading order in the $1/N$ expansion.
At zero temperature, the $1/N$ expansion was applied to $O(N)$ sigma models a
long time ago at leading order (LO) \cite{CJP} and at next-to-leading order
(NLO) \cite{Root}. 
At finite temperature the LO in $1/N$ has been studied in 
Ref.~\cite{M-O}. The effective potential in the large-$N$ limit is that of an 
ideal gas and thus straightforward to compute.
The NLO $1/N$ corrections to the free energy involve a momentum-dependent
self-energy and cannot be evaluated analytically.
In Ref.~\cite{jain}, the author therefore carried out a high-temperature
expansion to obtain purely analytical results for the LSM.
Similarly, in Ref.~\cite{kap} the authors
resorted to a ``high-energy'' approximation, which makes
the calculations manageable. However, this approximation is 
uncontrolled and it is difficult to assess how reasonable it is,
unless one calculates the full NLO $1/N$ corrections. 

The $O(N)$ sigma models have also been studied in detail at finite 
temperature using various other approaches. 
A systematic study has been carried 
out by Chiku and Hatsuda using optimized perturbation theory~\cite{Chiku}.
The method was used to calculate spectral functions, properties
of the effective potential, and dilepton emission rates. This method is
convenient from the point of view of renormalization. 
The Cornwall-Jackiw-Tomboulis or 2PI formalism~\cite{CJT} has also been used 
to examine various properties of the $O(N)$ linear sigma 
models at finite 
temperature~\cite{Baym,Amelino,Amelino2,Roh,Petro,Lenaghan,Nemoto}, 
see Ref.~\cite{Petro2} for a recent 
review. For example, in several papers the
temperature dependence of the pion and sigma masses, and of the vacuum 
expectation value of the sigma field, have been investigated. 
In the low-temperature phase,
the $O(N)$ symmetry is spontaneously broken and it is expected that
the symmetry is restored via a second-order phase transition.
The calculations of the effective potential as a function of
temperature have been carried out in the Hartree
approximation and the large-$N$
limit~\cite{Amelino,Amelino2,Roh,Petro,Lenaghan,Nemoto}. 
In these cases, the gap equations
for the propagators are easy to solve since the self-energy reduces to a
local mass term. In the Hartree
approximation, the result has been shown to be problematic (and a
first-order phase transition occurs), which has been remedied by
including more diagrams in the truncation~\cite{Verschelde,Baacke}, 
resulting in a second-order phase transition.
If one goes beyond the Hartree approximation or includes
the NLO contributions in the $1/N$ expansion,
the gap equations become nonlocal and very difficult to solve.

In this paper, we will study the 1PI effective potential at NLO in the
$1/N$ expansion without resorting to a high-temperature or high-energy 
approximation.
Thus the analysis presented here is an extension of the
papers by Jain~\cite{jain}, and by Bochkarev and Kapusta~\cite{kap}. 
We will follow the approach to the NLSM in 1+1 dimension of Ref.~\cite{ABW}. 
In the present case we do not need
to deal with thermal IR renormalons, which simplifies the
renormalization procedure. Nevertheless, we will obtain similar
conclusions about the renormalization of the effective potential in 3+1 
dimensions as in 1+1 dimensions. It turns out that at NLO, 
temperature-independent renormalization is only possible at the 
minimum of the effective potential. 
This aspect of the $1/N$ expansion was missed in previous work
\cite{jain,kap}, since the renormalization is considerably simplified 
or even ignored in the various approximations.

Since explicit chiral symmetry breaking plays a very important role in the 
actual 
hadron spectrum at low energy, we will consider also the case of explicit
symmetry breaking. The results change considerably and moreover, a critical 
temperature cannot be determined in that case, 
since the second-order phase transition turns into a smooth cross-over.  
 
The NLSM 
in 3+1 dimensions is nonrenormalizable and should
be viewed as an effective theory, which is valid up to a certain energy 
scale where new physics enters. Strictly speaking the LSM
is renormalizable,
but since it becomes a trivial theory in the limit where the cutoff goes to 
infinity, we treat it as a theory with a finite cutoff. 
Given a finite cutoff, we speak of divergences when
terms are increasing in magnitude without bound as the cutoff is 
increased. The low-energy physics
should be independent of such terms (decoupling) and one can 
subtract them in the renormalization procedure in order to 
avoid increasing 
sensitivity to the ultraviolet cutoff as it grows. On general grounds, one
expects the temperature dependence to be insensitive to an increasing 
cutoff due to the exponential suppression provided by the 
Bose-Einstein distribution. Therefore one expects 
the renormalization to be possible in a temperature-independent way. 

The paper is organized as follows. In Sec.~II, we discuss effective
actions of the LSM and NLSM in the $1/N$ expansion. 
In Sec.~III, we calculate the effective potential
and gap equations at NLO. In Sec.~IV, we discuss our results for 
the pressure at NLO for general $N$ 
and for the special case of $N=4$. Also, the so-called
high-energy approximation is discussed and compared with exact numerical 
results. In Sec.~V, we elaborate on 
the choice of parameters for $N=4$, in order to make contact 
with low-energy QCD phenomenology. In
Sec.~VI, we summarize and conclude.

\section{Effective actions}
The Euclidean Lagrangian of the $O(N)$-symmetric linear sigma
model with a symmetry breaking term $H$ is given by  
\bqa 
   \mathcal{L} &=& \frac{1}{2} \left(\partial_\mu 
  \phi_i\right)^2 + \frac{\lambda_b}{8 N} \left (
  \phi_i \phi_i\!-\!N f_{\pi,b}^2 \right)^2 - \sqrt{N} H \phi_N\;,
\label{eq:lagrangian_lin_sigma} 
\eqa 
where $i = 1 \ldots N$. Summation over repeated indices is implicitly
understood. The subscript $b$ denotes a bare quantity.
The coupling constants are rescaled with factors of $N$ 
in such a way that for large $N$ the action naturally scales as $N$. 

It is possible to eliminate the quartic interaction term from
Eq.~(\ref{eq:lagrangian_lin_sigma}) by introducing an auxiliary field
which is denoted by $\alpha$, in order to allow for Gaussian integration. 
To this end, we add to the Lagrangian 
Eq.~(\ref{eq:lagrangian_lin_sigma}) the term
\bqa
    \mathcal{L}_\alpha = \frac{N}{2 \lambda_b} \left[\alpha - 
    \frac{i \lambda_b}{2 N} \left(\phi_i \phi_i -  
    N f_{\pi,b}^2 \right) \right]^2 \;,
    \label{eq:lagrangian_alpha}
\eqa
such that one has
\bqa\nonumber
    \mathcal{L} &=& \frac{1}{2} \left( \partial_\mu \phi_i \right)^2 
 - \frac{i}{2} \alpha
  \left(\phi_i \phi_i - N f_{\pi,b}^2 \right) 
  \\ 
&&+ 
  \frac{N}{2 \lambda_b} \alpha^2
   -\sqrt{N} H \phi_N\;.
\eqa
Using the equation of motion for $\alpha$, one recovers the original
Lagrangian in Eq.\ (\ref{eq:lagrangian_lin_sigma}).
In the limit $\lambda_b\rightarrow\infty$, 
one obtains the Lagrangian of the
nonlinear sigma model. 

If explicit symmetry breaking is absent ($H=0$), 
the field $\phi$ acquires a vacuum expectation value by 
spontaneously breaking the symmetry. Because of the residual
$O(N-1)$ symmetry, we can write 
$\phi = (\pi_1, \pi_2, \ldots, \pi_{N-1}, \sigma)$, such that only
$\phi_N=\sigma$ has a nonzero expectation value.
For $H > 0$ the same argument applies, because the 
action is minimal when the $\sigma$ field is the only one that acquires an 
expectation value. 

Integrating over the $\pi$'s gives the following effective action
\bqa
  \nonumber
    \mathrm{S}_{\mathrm{eff}} &=&
{1\over2}(N-1)  
\mathrm{Tr} \log\left( -\partial^2 - i \alpha\right)  
   \\ && + 
        \int_{0}^{\beta} \! \mathrm{d}\tau\int \! \mathrm{d}^3 x \left[
       \frac{1}{2} \left( \partial_\mu \sigma \right)^2
  - \frac{i}{2} \alpha\, \sigma^2
  \right.  \nonumber \\  &&  \left.
         + \frac{i}{2} N f_{\pi,b}^2 \alpha 
         +  \frac{N}{2 \lambda_b} \alpha^2 
      - \sqrt{N} H \sigma   
   \right] \;.
\label{sigalp}
\eqa
We next parametrize the quantum fields $\sigma$ and $\alpha$ by
writing them as a sum of space-time independent vacuum expectation
values $im^2$ and $\bar{\sigma}$, 
and quantum fluctuating fields $\tilde{\alpha}$ and $\tilde{\sigma}$: 
\bqa
\alpha&=&im^2+\frac{\tilde{\alpha}}{\sqrt{N}}\;,
\label{s1}\\
\sigma&=&\sqrt{N}\bar{\sigma}+\tilde{\sigma}
\label{s2}
\;.
\eqa
Using
Eq.~(\ref{eq:lagrangian_alpha}) one can show that the vacuum
expectation value of $\alpha$ is purely imaginary. The vacuum
expectation value of $\sigma$ is proportional to $\sqrt{N}$, which
follows from Eq.~(\ref{eq:lagrangian_lin_sigma}). 
Substituting Eqs.~(\ref{s1}) and~(\ref{s2}) into
Eq.~(\ref{sigalp}), the effective action $S_{\rm eff}$ can be written as
\bqa\nonumber
    \mathrm{S}_{\mathrm{eff}} &=& 
{1\over2}(N-1)\mathrm{Tr} 
       \log\left( -\partial^2 + m^2 - \frac{i \tilde \alpha}{\sqrt{N}}\right) 
\\ && \nonumber
        - \beta V N H \bar \sigma  
+  \int_{0}^{\beta}\mathrm{d} \tau \int
\mathrm{d}^3x 
   \bigg[
       \frac{1}{2} \left ( \partial_\mu \tilde \sigma \right)^2
\\ && \nonumber
     + \frac{1}{2} \left(m^2\!-\! 
        \frac{i \tilde \alpha}{\sqrt{N}}\right) 
        \left(\sqrt{N} \bar \sigma\!+\!\tilde \sigma\right)^2
    \\ &&  \nonumber 
        -  \frac{N}{2} f_{\pi,b}^2 
\left(m^2\!-\!\frac{i \tilde \alpha}{\sqrt{N}} 
        \right) - \frac{N}{2 \lambda_b} 
        \left(m^2\!-\!\frac{i \tilde \alpha}{\sqrt{N}}\right)^2 
    \\ &&  
 - \sqrt{N} H \tilde \sigma
\bigg] \;.
  \label{eq:shiftedeffaction}
\eqa
Expanding Eq.~(\ref{eq:shiftedeffaction}) in powers of $1/\sqrt{N}$ up to
corrections of order $1/\sqrt{N}$, 
one finds~\footnote{The terms that are linear
in $\tilde \alpha$ and $\tilde \sigma$ vanish at the 
minimum of the effective potential.}
\begin{multline}
    \frac{\mathrm{S}_{\mathrm{eff}}}{\beta V} = 
{1\over2}(N-1)
\sumint_P \log \left (P^2 + m^2 \right)
        - \frac{N m^2}{2} \left (f^2_{\pi, b} - \bar \sigma^2 \right)
\\
        - \frac{N m^4}{2 \lambda_b}
        - N H \bar \sigma  + \sqrt{N} \times \mathrm{terms\;linear\;in\;}
\tilde \alpha\;\mathrm{and}\;\tilde \sigma 
       \\  +  \frac{1}{2} \sumint_P 
\chi^\mathrm{T}
\left(
\begin{array}{cc} 
\tfrac{1}{2} \Pi(P, m) + \frac{1}{\lambda_b}  
  & -i \bar \sigma \\
  -i \bar \sigma & P^2 + m^2  
\end{array}
\right)
\chi^*
  \label{eq:expandedeffaction} \;,
\end{multline}
where $\chi^T = (\tilde \alpha_P$,\, $\tilde \sigma_P)$ is a vector
containing the Fourier transforms of $\tilde \alpha$ and $\tilde
\sigma$, and the function $\Pi(P,m)$ is given by
\bqa
 \Pi(P, m) &=& \sumint_Q \frac{1}{Q^2+m^2} \frac{1}{(P+Q)^2+m^2} \;.  
\eqa
We have introduced the sum-integral
\bqa
  \sumint_Q&\equiv&T\sum_{q_0 = 2\pi nT}\int 
  \frac{\mathrm{d}^3q}{(2\pi)^3}\;,
\eqa
which involves a summation over Matsubara frequencies $q_0$ and an integral
over three-momenta $q$. For later convenience we also introduce a symbol for
the difference between a sum-integral and an integral 
\bqa
\sumdiff_Q&\equiv&\sumint_Q-\int_Q\;,
\eqa
where $\int_Q = \int \mathrm{d}^4Q/(2\pi)^4$.

\section{Effective potential and gap equations}

One can obtain the effective potential through next-to-leading order
in the $1/N$ expansion from Eq.~(\ref{eq:expandedeffaction})
by performing the Gaussian integral over the fluctuating fields $\tilde
\alpha$ and $\tilde \sigma$. 
Up to corrections of order $1/N$, the effective potential can be written
as
\bqa
{\cal V}(m^2, \bar \sigma)&=&
  N \mathcal{V}_{\mathrm{LO}}(m^2,\bar \sigma) 
+ \mathcal{V}_{\mathrm{NLO}}(m^2, \bar \sigma)\;,
\label{nloeq}
\eqa
where
\begin{eqnarray}\nonumber
  \mathcal{V}_{\mathrm{LO}}\left(m^2, \bar \sigma \right)
 &=& \frac{m^2}{2} \left ( f_{\pi,b}^2 - \bar \sigma^2 \right) + 
        \frac{m^4}{2 \lambda_b} + H \bar \sigma
\\
  &&  -\frac{1}{2} \sumint_P \log(P^2 + m^2 )  
  \label{eq:effpotlo}
   \;, \\ 
   \mathcal{V}_{\mathrm{NLO}}\left(m^2, \bar \sigma\right) 
  &=& -\frac{1}{2} \sumint_P \log I(P,m) 
\;.
\end{eqnarray}
Here,
\beq
  I(P,m) = 16 \pi^2 \Pi(P, m) + \frac{32 \pi^2}{\lambda_b} 
+ \frac{32 \pi^2 \bar \sigma^2}
      {P^2 + m^2} \;.
\label{Ipm}
\eeq
To derive the effective potential, we subtracted divergent constants
which are independent of $\bar\sigma$, $m$ and the temperature. 
Equivalently, these terms can be removed by adding a vacuum counterterm
$\Delta{\cal E}$ to the effective potential.
In the following, we simply drop these terms.

In thermodynamic equilibrium, the system will be in
the state that minimizes the effective potential with respect to $m^2$
and maximizes it with respect to $\bar \sigma$. This difference is due
to the fact that the vacuum expectation value of $\alpha$ is
imaginary and that of $\sigma$ is real. 
Differentiating the NLO effective potential with respect to $m^2$
and $\bar{\sigma}$, one obtains
\bqa
  \sumint_P \frac{1}{P^2 + m^2}
   - \frac{2 m^2}{\lambda_b} +
  \frac{1}{N} \sumint_P 
   \frac{ \frac{\mathrm{d} \Pi(P, m)}{\mathrm{d} m^2}
  - \frac{2 \bar \sigma^2}{ \left( P^2 + m^2 \right)^2}}
  {\Pi(P, m) + \frac{2}{\lambda_b}+ \frac{2\bar \sigma^2}
      {P^2 + m^2}} \nonumber \\
  =
   \left( f_{\pi,b}^2 - \bar \sigma^2\right) \;, 
   \hspace{1cm}
  \label{eq:gapmsq} 
  \\
  \left(m^2 + \frac{2}{N} \sumint_P \frac{1}{P^2 + m^2} 
  \frac{1}{\Pi(P,m) + \frac{2}{\lambda_b} + 
  \frac{2 \bar \sigma^2}{P^2 + m^2}} \right)\bar \sigma 
  \nonumber \\
   = H \;. \hspace{1cm}
  \label{eq:gapsigma}
\eqa
These equations are often referred to as gap equations.
Solving the gap equations gives $m$ and $\bar \sigma$ as a
function of the parameters $f_\pi$, $H$ and $\lambda$, and of the temperature. 

The inverse $\tilde \alpha$ and $\tilde \sigma$ propagators can be
obtained from Eq.~(\ref{eq:expandedeffaction}). One finds
\begin{eqnarray}
  D^{-1}_{\tilde \alpha} (P, m)  &=& \frac{1}{2} \Pi(P,m) + 
  \frac{1}{\lambda_b} + \frac{\bar \sigma^2}{P^2 + m^2} \;, \\
  D^{-1}_{\tilde \sigma} (P, m)  &=& P^2 + m^2 + 
  \frac{2 \bar \sigma^2}{\Pi(P, m) + 2 / \lambda_b}\;.
\label{inversesigmaprop}
\end{eqnarray}
The values for $m^2$ and $\bar \sigma$ are determined by solving the
gap equations. Using the $\tilde \alpha$ propagator, one finds
that the inverse $\pi$ propagator is given by
\begin{multline}
 D^{-1}_{\pi} (P, m) = P^2 + m^2 \\ + 
  \frac{2}{N} \sumint_Q  \frac{1}{(P+Q)^2 + m^2} 
  \frac{1}{\Pi(Q, m) +
\frac{2}{\lambda_b} + \frac{2 \bar \sigma^2}{Q^2 + m^2}} \;.
\end{multline}
From this equation and the gap equation (\ref{eq:gapsigma}) 
one sees that also at NLO
in the broken phase where $\bar \sigma \neq 0$ (for $H=0$), the pions are 
massless, in accordance with Goldstone's theorem. 

From Eq.\ (\ref{inversesigmaprop}) one can see that in the unbroken phase, 
the $\sigma$ mass becomes equal to the LO mass of the $\pi$ field, which is 
$m^2$. It may appear therefore that the $\sigma$ and $\pi$ masses are not 
equal at next-to-leading order, but this is not a correct conclusion.
We note that the $\sigma$ field only starts to propagate at NLO, so its 
$1/N$ mass corrections require a next-to-next-to-leading order calculation. 
We emphasize that in the calculation of the NLO pressure one only needs the 
LO masses, as we will see explicitly below in Eq.\ (\ref{gapexp}). 

In subsecs.~\ref{sub1} and~\ref{sub2}, 
we will explicitly calculate the leading-order 
and next-to-leading order contributions to the effective
potential in $3+1$ dimensions. We will evaluate integrals using an
ultraviolet momentum cutoff $\Lambda$ and assume that $\Lambda \gg m,
2 \pi T$.

\subsection{Leading-order contribution}
\label{sub1}
The leading-order contribution to the effective potential is
\bqa\nonumber
{\cal V}_{\rm LO}&=& 
  \frac{m^2}{2} \left(f^2_{\pi,b}- \frac{\Lambda^2}{16\pi^2} - 
  \bar{\sigma}^2\right)
+ \frac{T^4}{64\pi^2} J_0(\beta m)
\\&& +
\frac{m^4}{64\pi^2}\left[
  \frac{32 \pi^2} {\lambda_b} + 
  \log \left ( \frac{\Lambda^2}{m^2} \right)
+ \frac{1}{2}  
\right] + H \bar \sigma\;.
\label{LO}
\eqa
Eq.\ (\ref{LO}) contains ultraviolet divergences 
in the sense explained in the introduction.
These divergences can be dealt with by defining the renormalized
parameters $f_{\pi}^2$ and $\lambda$ as
\bqa
f^2_\pi &=& f^2_{\pi,b} -
\Lambda^2 / 16 \pi^2
\\ 
  \frac{32\pi^2}{\lambda} &= &\log
  \left(\frac{\Lambda^2}{\mu^2} \right) + \frac{32 \pi^2}{\lambda_b} \;,
  \label{eq:lambdarenorm}
\eqa
where $\lambda = \lambda(\mu)$. 
After this renormalization, the
leading-order effective potential becomes
\bqa
  \mathcal{V}_{\mathrm{LO}}
  &=& \frac{m^2}{2}\left(
  f_{\pi}^2 - \bar \sigma^2 \right) 
    +\frac{m^4}{64 \pi^2} 
  \left[\frac{32 \pi^2}{\lambda} + 
  \log \left ( \frac{\mu^2}{m^2} \right)
   + \frac{1}{2} \right] \nonumber \\
  &&+ 
  \frac{T^4}{64 \pi^2} J_0(\beta m) + H \bar \sigma\;,
  \label{eq:effpotlo3d}
\eqa
where the function $J_0(\beta m)$ is
\begin{equation}
  J_0(\beta m) = \frac{32}{3 T^4} \int_0^{\infty}\mathrm{d}p\;\frac{p^4}{\omega_p}
  n(\omega_p)\;.
\label{Jzero}
\end{equation}
Here, $n(\omega_p) = [\exp(\beta \omega_p) - 1]^{-1}$ is the Bose-Einstein
distribution function. Note that one makes an error in the
evaluation of $J_0$ by integrating up to infinite momenta 
instead of up to $\Lambda$. However, this error is negligible as long as 
$\Lambda \gg m, 2 \pi T$. This remark 
also applies to the functions $J_1$, $K_0^\pm$ and 
$K_1^\pm$ defined below. For an investigation of how to deal with a finite cut-off
in the calculation of sum-integrals cf.\ Ref.\ \cite{Amte}. 

The renormalization group equation for the running coupling $\lambda$
that follows from Eq.~(\ref{eq:lambdarenorm}) is
\begin{equation} 
  \beta(\lambda) = \mu \frac{d
  \lambda}{d \mu} = \frac{\lambda^2}{16 \pi^2} \;.
\label{beta1}
\end{equation} 
Note that the $\beta$-function is exact to all orders in
$\lambda^2$ in the large-$N$
limit, but differs from 
the perturbative one obtained at one loop. However, at NLO they agree.

Since the potential term in the Lagrangian should always have
a minimum in order to have a stable theory, $\lambda_b$ must be positive (cf.\
\cite{Amelino2} for a detailed discussion). 
From Eq.\ (\ref{eq:lambdarenorm}) it immediately 
follows that there is a maximal value for the cutoff given by
\begin{equation}
  \Lambda_\mathrm{max} = \mu \exp \left( \frac{16 \pi^2}{\lambda}
  \right) \;.
\label{cutmax}
\end{equation}
Therefore this theory should be viewed as an effective theory, which
is valid up to the cutoff given by Eq.~(\ref{cutmax}).  Taking the cutoff to
infinity is equivalent to taking $\lambda$ to zero, which implies 
that the theory is trivial. 
One should keep in mind that the renormalized leading-order
effective potential does not depend explicitly on $\Lambda$, but is
only valid for $m$ and $T$ much smaller than $\Lambda_{\mathrm{max}}$.
When $\Lambda = \Lambda_{\mathrm{max}}$ the linear sigma model reduces
to the nonlinear sigma model, since in this case $\lambda_b = \infty$.

The leading-order renormalized gap equations 
follow from differentiating Eq.\ (\ref{eq:effpotlo3d}) with respect to
$m^2$ and $\bar{\sigma}$ and are given by
\begin{eqnarray}
  G & = & 16 \pi^2 f^2_\pi \;,
 \label{eq:logapeqm}
  \\
  H &=& m^2 \bar \sigma  \;,
 \label{eq:logapeqsigma}
\end{eqnarray}
where
\beq
G  =
T^2 J_1(\beta m) + 16 \pi^2 \bar \sigma^2 
  - m^2 \log \left( \frac{\mu^2}{m^2} \right) 
    -  \frac{32\pi^2 m^2}{\lambda} 
\;.
\eeq
Here, we have defined the function $J_1(\beta m)$ as
\begin{equation}
 J_1(\beta m) =  \frac{8}{T^2} \int_0^{\infty}
\mathrm{d}p\;\frac{p^2}{\omega_p} n(\omega_p) \;.
\label{J1}
\end{equation}
If $H=0$, one can show by using the gap equation (\ref{eq:logapeqsigma}) 
that either $m=0$ or $\bar \sigma = 0$. From the gap equation
(\ref{eq:logapeqm}) it follows that for $m=0$ the 
expectation value of $\sigma$ has the temperature dependence
\begin{equation}
  \bar \sigma = \sqrt{f_\pi^2 - \frac{T^2}{12}}\;.
\end{equation}
At $T = T_c \equiv \sqrt{12} f_\pi$ there is a second-order phase
transition \cite{kap}. Below $T_c$ the $O(N)$ symmetry is broken 
spontaneously to $O(N-1)$ since $\bar
\sigma \neq 0$. Above $T_c$ the $O(N)$ symmetry is restored and one
has $\bar \sigma = 0$ and $m \neq 0$.

\subsection{Next-to-leading order contribution}
\label{sub2}
In this section we will show that it is not possible to renormalize
the next-to-leading order effective potential in a 
temperature-independent way. It turns out that we can only renormalize the
effective potential at the minimum, since the temperature-dependent
divergences become temperature independent by using the leading order gap
equations. To show this, we will extract
the divergent parts of the effective potential, which can be done
analytically. 

In order to isolate all divergences, in principle we need to evaluate 
$\Pi(P,m)$ including corrections of order $m^2 / \Lambda^2$, since 
such terms can also give
rise to divergences in the effective potential. However, since the
linear sigma model is an effective theory, 
Eq.~(\ref{eq:lagrangian_lin_sigma}) should 
be viewed as the part containing only
the
relevant operators. For instance, we have not included irrelevant operators of
dimension six, which also contribute to $\Pi(P,m)$ at order
$1/\Lambda^2$. Therefore, for consistency with 
Eq.~(\ref{eq:lagrangian_lin_sigma}) we do not consider order 
$1/\Lambda^2$ terms in $\Pi(P,m)$~\cite{Polchinski}. 
We only retain the unsuppressed terms of
$\Pi(P,m)$ in an expansion in $1/\Lambda^2$, since this expression is
much less complicated than the exact one. We find
\begin{multline}
\Pi(P,m)
= \frac{1}{16 \pi^2} \left[ \log \left( \frac{\Lambda^2}{m^2} \right)
  + 1 
  \right. \\ \left. 
  + \sqrt{\frac{P^2+4m^2}{P^2}} 
  \log \left( \frac{\sqrt{P^2+4m^2} - \sqrt{P^2}}{\sqrt{P^2+4m^2} + 
  \sqrt{P^2}} \right) \right] \\ 
+ \Pi_T(P,m) \;,
\end{multline}
where the temperature-dependent part of $\Pi(P,m)$ equals
\begin{multline}
\Pi_T(P,m) = \\
   \frac{1}{8\pi^2p}\int_0^{\infty}dq\;\frac{q}{\omega_q}
  \log \left( \frac{q^2 + p q + A^2}{q^2 - p q + A^2} \right)
  n(\omega_q) \;.
\end{multline}
Here
\begin{equation}
  A^2 = \frac{P^4 + 4 m^2 p_0^2}{4 P^2} \;.
\label{Akwadraat}
\end{equation}
In the limit $P \gg m, T$, we can approximate $\Pi_T(P,m)$ by
\begin{multline}
  \Pi_T(P,m) \approx 
 \frac{1}{8 \pi^2} \left[
  \frac{T^2}{P^2} J_1(\beta m) - 
  \frac{4m^2 T^2 p_0^2}{P^6} J_1(\beta m) 
  \right. \\ - \left.
  \frac{\left(3 P^2 - 4 p^2\right)T^4}{P^6} J_0(\beta m)
  \right] \;.
\end{multline}

The next-to-leading-order effective potential has only
ultraviolet divergences. Using the leading order renormalization of 
$\lambda_b$, it is easily seen that $I(P,m)$ becomes finite. 
Also, the difference 
\beq
\sumdiff_P \log I(P,m) 
\;,
\eeq
is finite (cf.\ Sec.~\ref{4b}). 
Therefore, all possible divergences
of $\mathcal{V}_{\mathrm{NLO}}$ can be isolated by calculating
\beq
   -\frac{1}{2} \int_P \log I_{\mathrm{HE}}(P, m) \;,
\eeq
where $I_{\mathrm{HE}}(P,m)$ is the high-energy (HE) approximation
to $I(P,m)$. It gives the large-$P$ behavior of $I(P,m)$.
After averaging over angles, we find 
\begin{multline}
  \log I_{\mathrm{HE}} =
  \log C_1 + \frac{1}{P^2} \frac{C_2}{C_1} 
- \frac{1}{2 P^4} 
  \left( \frac{C_2}{C_1}
  \right)^2 
  + \frac{1}{P^4} \frac{C_3}{C_1} \;,
\end{multline}
where
\bqa
  C_1 &=& \log \left( \frac{\mu^2}{P^2} \right)+ 1 +
   \frac{32\pi^2}{\lambda} \;,
\\
  C_2 &=& - 2 m^2 \left[1 + \log \left(\frac{P^2}{m^2} \right) \right]
 \nonumber \\
        &&   + 32\pi^2 \bar \sigma^2 + 2 T^2 J_1(\beta m)\;,
\\
  C_3 &=& 
+ m^4 \left[2 \log \left( \frac{P^2}{m^2} \right) - 1 \right]
\nonumber \\
 &&   - m^2 \left[ 32 \pi^2 \bar \sigma^2 + 2 T^2 J_1(\beta m) \right] \;.
\eqa
By integrating the function $\log I_{\mathrm{HE}}$ over $P$, we obtain
all the divergences of the NLO effective potential.  The
logarithmic and power divergences are given by the quantity $D$, which
is 
\begin{eqnarray}
  D &=&
\frac{1}{16\pi^2} 
\biggl \{ \biggr. 
   \Lambda^2  e^{1+32\pi^2 / \lambda_b} 
  \,\mathrm{li} \left( \frac{1}{e^{1+32\pi^2 / \lambda_b}} \right) G 
 \nonumber \\ 
   & & 
  \phantom{ \frac{1}{16\pi^2} 
\biggl \{ \biggr.}
  - m^2 \Lambda^2 \left[1 +
 2 e^{1+32\pi^2/\lambda_b}
  \, \mathrm{li} 
  \left( \frac{1}{e^{1+32\pi^2 / \lambda_b}} \right) 
  \right]
 \nonumber \\
  && 
   \phantom{ \frac{1}{16\pi^2} 
\biggl \{ \biggr.}
  + 2 m^4 \log \left(\frac{\Lambda^2}{m^2} \right) 
  \biggl. \biggr \} \;,
\label{eq:nlodivs} 
\end{eqnarray}
while the terms that have a small cutoff dependence through their
dependence on $\lambda_b$, are given by the quantity $E$, which is
\begin{eqnarray}
E &=& 
  \frac{1}{16\pi^2} 
  \Biggl [ \Biggr. 
  3 m^2 \left(-G + \tfrac{3}{2} m^2 \right) \log
   \left(
         1 + \frac{32\pi^2}{\lambda_b} \right)
 \nonumber \\ 
  & & + \left(G - 2 m^2 \right)^2 \frac{1}{ 1 + \frac{32\pi^2}{\lambda_b}}
  \Biggl. \Biggr ] \;.
  \label{eq:slowdiv}
\end{eqnarray}
Since $G$ depends explicitly on the temperature, it is impossible to
renormalize the next-to-leading-order effective potential in a 
temperature-independent way. However, at
the minimum, one can use the leading-order gap
equation~(\ref{eq:logapeqm}), to show that $G = 16\pi^2
f_\pi^2$. Hence, the divergences become independent of the temperature 
at the
minimum and we can renormalize in a temperature-independent manner.
We discuss this next.

The divergence in the first line of Eq.~(\ref{eq:nlodivs}) is
independent of $m$ in the minimum.  This divergence can be removed by
vacuum renormalization.
The divergent terms which are proportional to $m^2$
can be removed by defining the renormalized parameter $f_\pi$ as
\begin{equation}
\begin{split}
  f_\pi^2 = &f^2_{\pi, b} - \left(1 + \frac{2}{N} \right)
  \frac{\Lambda^2}{16\pi^2} 
\\
   &-
  \frac{1}{N} \frac{\Lambda^2}{4\pi^2}
  \bigg[ 
  e^{1+32\pi^2/\lambda_b} \, \mathrm{li}  
 \left( \frac{1}{e^{1+32\pi^2 / \lambda_b}} \right) \bigg]
  \;.
  \label{eq:fpirenormnlo}
\end{split}
\end{equation}
The remaining divergence is proportional to $m^4$
and is removed by renormalizing
$\lambda_b$ as follows
\begin{equation} 
  \frac{32\pi^2}{\lambda} = 
   \frac{32 \pi^2}{\lambda_b} + 
\left(1 + \frac{8}{N} \right)
 \log
\left(\frac{\Lambda^2}{\mu^2} \right)
  \label{eq:lambdarenormnlo} \;.
\end{equation} 
From Eq.~(\ref{eq:lambdarenormnlo}), we obtain the $\beta$-function 
governing the running of $\lambda$:
\begin{equation}
\beta(\lambda)
= \frac{\lambda^2}{16 \pi^2} \left( 1+ \frac{8}{N} \right)\;,
\end{equation}
which coincides with the standard one-loop $\beta$-function 
in perturbation theory.
One could argue from the renormalization
that one can only trust the $1/N$ expansion for $N \gg 8$. 
Although the $1/N$
correction to the $\beta$ function indeed has a large coefficient, this is not
the case for the effective potential itself as we will see below. 

As mentioned, the terms in $E$ have a small cutoff dependence through their
dependence on $\lambda_b$. We do not renormalize them, since they do not grow
without bound with increasing cutoff and are not strictly speaking divergences.
The effective potential does not become increasingly sensitive to them
with increasing cutoff \footnote{Since $\Lambda \leq \Lambda_\mathrm{max}$, 
one also has to increase $\Lambda_\mathrm{max}$ accordingly, by considering 
decreasing values of $\lambda$.}. 
The term in the first
line of Eq.~(\ref{eq:slowdiv}) becomes smaller if we increase $\Lambda$ and
the absolute value
of the other term from $E$ increases as a function of $\Lambda$, but
is bounded by a finite number which is independent of $\lambda_b$ and 
$\Lambda$. Moreover, 
renormalizing these terms would invalidate the $1/N$ expansion, because of their
magnitude. This is similar to ordinary perturbation theory,
where 
one is only allowed to do finite renormalizations that do not invalidate the
perturbative expansion.  
A final reason for not renormalizing these terms is the connection
with the nonlinear sigma model ($\lambda_b = \infty$). In that case, the 
terms from $E$ are not divergent and we would choose 
to renormalize $f_\pi$ just as in Eq.~(\ref{eq:fpirenormnlo}) with
$\lambda_b = \infty$.

The NLO correction changes the critical temperature $T_c$. Since the
NLO gap equations are complicated, we are not able to get an
analytical expression for the NLO critical temperature. In the limit of
small $\lambda_b$ and $H=0$ the gap equations however simplify to
\bqa
 && \sumint_P \frac{1}{P^2 + m^2}
   - \frac{2 m^2}{\lambda_b} 
  =
   \left( f_{\pi,b}^2 - \bar \sigma^2\right) \;, 
   \hspace{1cm}
  \label{eq:gapmsqsmalllambda} 
  \\
 && \left(m^2 + \frac{\lambda_b}{N} \sumint_P \frac{1}{P^2 + m^2}  
  \right) \bar \sigma
   = 0 \;.
  \label{eq:gapsigmasmalllambda}
\eqa
From the gap equations it follows that the critical temperature at NLO is 
\beq
T_c = \sqrt{\frac{12}{1+2/N}} f_\pi \;.
\eeq
This result is the same as obtained in Refs.~\cite{kap, jain}. 
This result is probably 
only correct in the weak-coupling limit and $T_c$ 
may depend on $\lambda$ at NLO in $1/N$. 
As we will see in the next section the transition is of second (or
higher) order.

\section{Pressure}
The pressure ${\cal P}(T)$ is equal to the value of the effective
potential at the minimum at temperature $T$ minus its value at the
minimum at zero temperature. As we showed in the previous section, we
can renormalize the effective potential at the minimum. The pressure
is therefore a well-defined quantity.  In order to determine the NLO
effective potential in the minimum, we need the gap equation only to
leading order~\cite{Root}. Writing the solutions to the gap equations as 
\bqa
m^2 &=
&m^2_\mathrm{LO} + m^2_{\mathrm{NLO}} / N\\
\bar \sigma &=& \bar \sigma_{\mathrm{LO}} + \bar \sigma_{\mathrm{NLO}}
/ N \;,
\eqa 
and Taylor expanding the effective
potential~(\ref{nloeq}), we obtain (up to ${\cal O}(1/N)$ corrections)
\bqa\nonumber
  \mathcal{V}(m^2, \bar \sigma) &=& 
  N \mathcal{V}_{\mathrm{LO}}(m^2_{\mathrm{LO}}, \bar
  \sigma_\mathrm{LO}) 
   + \mathcal{V}_{\mathrm{NLO}}(m_{\mathrm{LO}}^2, \bar
\sigma_{\mathrm{LO}})  
  \\ && \nonumber
  + m^2_{\mathrm{NLO}}
  \left. \frac{\partial \mathcal{V}_{\mathrm{LO}}(m^2)} 
    {\partial m^2} \right \vert_{m^2 = m^2_\mathrm{LO}} 
\\ &&
+  \bar \sigma_{\mathrm{NLO}}
  \left. \frac{\partial \mathcal{V}_{\mathrm{LO}}(\bar \sigma)} 
    {\partial  \bar \sigma} \right \vert_{\bar \sigma = \bar
  \sigma_\mathrm{LO}}\;.
\label{gapexp}
\eqa
The last two lines of Eq.~(\ref{gapexp}) vanish by using the leading-order 
gap equations. 
In the following, we will 
write the pressure ${\cal P}$ as
\bqa
\mathcal{P} \equiv N
\mathcal{P}_{\mathrm{LO}} + \mathcal{P}_{\mathrm{NLO}} 
\;.
\eqa
From the discussion above, it follows that
\begin{eqnarray}
  \mathcal{P}_{\mathrm{LO}}  &=& 
     \mathcal{V}^{T}_{\mathrm{LO}}(m^2_T, 
\bar \sigma_T) 
   - \mathcal{V}^{T=0}_{\mathrm{LO}}(m^2_0, 
\bar \sigma_{0})  
  \;, \label{eq:lopres}
  \\
  \mathcal{P}_{\mathrm{NLO}} &=&
     \mathcal{V}^{T}_{\mathrm{NLO}}(m^2_T, 
  \bar \sigma_T) 
   - \mathcal{V}^{T=0}_{\mathrm{NLO}}(m^2_0, 
\bar \sigma_0) \;,
\end{eqnarray}
where $m^2_T$ and $\bar \sigma_T$ are the solutions of the leading-order 
gap equations~(\ref{eq:logapeqm}) 
and~(\ref{eq:logapeqsigma}), at temperature $T$. 

In the following, 
we will present the results for the numerical
evaluation of the leading and the next-to-leading order contributions
to the pressure for general $N$. In subsecs.~\ref{4c} and~\ref{4d}, we
specialize to $N=4$. As we will motivate in Sec.~\ref{sec:parameters}, 
we will use the
following values for the parameters: $\lambda (\mu =
100\;\mathrm{MeV}) = 30$, $f_\pi = 47\;\mathrm{MeV}$ (note that our $f_\pi$ 
is $1/2$ times the more conventional definition) and if there is explicit
symmetry breaking $H = \left(104 \;\mathrm{MeV} \right)^3$.
Otherwise $H$ vanishes.
A realistic choice of parameters would allow us to compare to
lattice-QCD simulations of the $N_f=2$ case for $T \simordertwo T_c$, although
this would require extrapolation of the lattice results down to the actual
pion masses.  

\subsection{Leading-order contribution to the pressure}
\label{4a}
\begin{figure}[htb]
\begin{center}
\scalebox{0.9}{\includegraphics{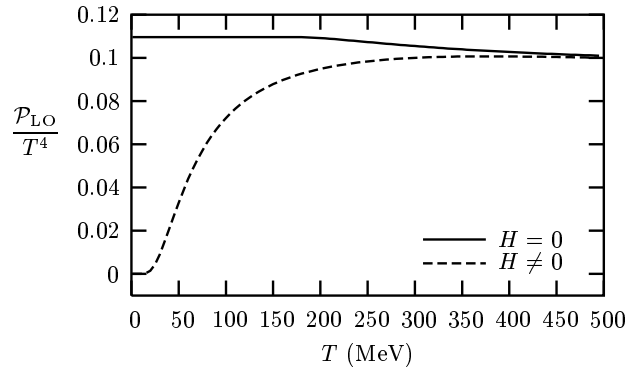}}
\caption{Leading-order pressure
${\cal P}_{\rm LO}$ normalized to $T^4$ as a function of
temperature without and with explicit symmetry breaking.}
\label{fig:leadingpressure}
\end{center}
\end{figure}

Using Eqs.~(\ref{eq:effpotlo3d}) and (\ref{eq:lopres}), we can easily
calculate the leading-order pressure. 
In Fig.~\ref{fig:leadingpressure}, we show 
the leading-order pressure normalized by 
$T^4$. If $H=0$, the pions are massless below $T_c$ and 
the leading-order effective potential~(\ref{eq:effpotlo3d})
reduces to the free energy of an ideal gas of massless particles:
${\cal V}_{\rm LO}=\pi^2 T^4 / 90$.
\subsection{Next-to-leading order contribution to the pressure} 
\label{4b}
To calculate the next-to-leading order contribution to the pressure, we
decompose $\mathcal{P}_{\mathrm{NLO}}$ as follows
\begin{equation}
  \mathcal{P}_{\mathrm{NLO}} = D(m_T) - D(m_0) + F_1 + F_2 \;,
  \label{eq:nlopressplit}
\end{equation}
where $D(m)$ is the term containing logarithmic and power ultraviolet
divergences given in Eq.~(\ref{eq:nlodivs}),
and $F_1$ and $F_2$ are finite terms defined below. 

The term $F_1$ has a weak cutoff
dependence and is defined by
\begin{eqnarray}
  F_1 &=& -\frac{1}{2} \int_P \Bigg\{ \log \left [\Pi(P, m_T) + \frac{2}{\lambda_b} +
   \frac{2\bar \sigma_T^2}{P^2 + m_T^2}
 \right] 
  \nonumber \\  
 &&\phantom{ -\frac{1}{2} \int_P }
   -\log \left [\Pi(P, m_0) + \frac{2}{\lambda_b} +
   \frac{2\bar \sigma_0^2}{P^2 + m_0^2}
 \right ]  \Bigg\}
 \nonumber \\
 && - D(m_T) + D(m_0) \;. 
  \label{eq:f1}
\end{eqnarray}
We calculated $F_1$ numerically by rewriting the terms involving $D$ as an
integral. We can then subtract the integrands, instead of
the large values of the integral. In this way, it is easier to avoid large
numerical errors.

The function $F_2$ is defined by
\bqa\nonumber
   F_2&=&- \frac{1}{2} 
\sumdiff_P \log \left [\Pi(P, m_T) + \frac{2}{\lambda_b} +
   \frac{2\bar \sigma_T^2}
      {P^2 + m_T^2} \right] \;.
 \label{eq:f2}
\eqa
In order to calculate the function $F_2$, we have modified 
the Abel-Plana formula~\cite{Barton} to obtain the relation
\bqa\nonumber
  \Delta_N &=&\sum_{n = N}^{\infty} f(p_0 = 2 \pi n) -
  \frac{1}{2\pi} \int^{\infty}_{a} \mathrm{d} p_0 f(p_0) 
\\ &=&
  \frac{1}{\pi} \int_0^{\infty} \mathrm{d}\rho \; \mathrm{Im} f(a + i
  \rho) \frac{1}{e^\rho  + 1} \;,
\eqa
where $a = 2 \pi (N - 1/2)$. This formula is valid as long as $f(p_0)$
has no poles or cuts for $\mathrm{Re}(p_0) \geq a$, $f(p_0) \in \mathbb{R}$ 
for $p_0
\in \mathbb{R}$ and $f(p_0)$ grows slower than an exponential for $p_0
\rightarrow \infty$. This relation is useful for numerical
calculations, since it prevents us from subtracting two large
quantities.
In our case $f(p_0) = \log I(P,m)$ which is even in $p_0$
and has a cut for $\mathrm{Re}(p_0) = 0$. Then we can use that 
\beq
  \Delta_{-\infty} = 2 \Delta_N + \sum_{n = -N+1}^{N-1} f(p_0 = 2 \pi n) -
  \frac{1}{2\pi} \int^{a}_{-a} \mathrm{d} p_0 f(p_0) \;,
\eeq
where $N \geq 1$ because of the cut. We checked that changing $N$ has
no effects on the 
results. After calculating $\Delta_{-\infty}$, we integrate over
three-momentum $p$ up to $\Lambda_\mathrm{\max}$, 
which gives a finite result for $F_2$. We observe that 
the difference of a sum-integral and an integral is dominated by the 
low-momentum modes. 
This shows that the high-energy approximation in Ref.~\cite{kap} 
applied to their ``interaction pressure'' 
(implicitly defined as such a difference and directly related to $F_2$)
is invalid, since
the high-momentum modes are assumed to give the main 
contribution. 

\begin{figure}[htb]
\begin{center}
\scalebox{0.9}{\includegraphics{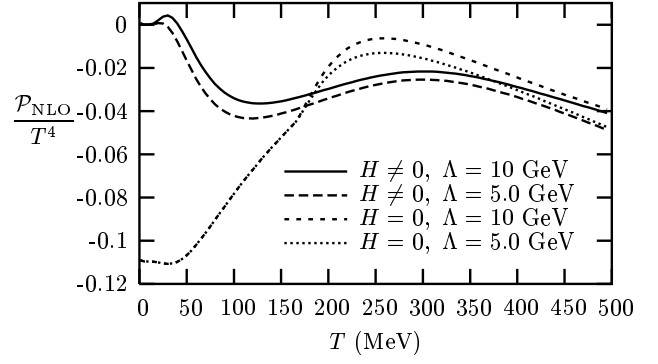}}
\caption{Next-to-leading order contribution to the pressure 
normalized to $T^4$, as function
of temperature for $H=0$ and $H = \left(104 \;\mathrm{MeV} \right)^3$,
for different values of the
cutoff $\Lambda$. }
\label{fig:nlopressure}
\end{center}
\end{figure}

After renormalization, we
find that the next-to-leading order contribution to the pressure is 
\begin{equation}
  \mathcal{P}_{\mathrm{NLO}} = \frac{m_T^4}{8\pi^2} \log
  \left(\frac{\mu^2}{m_T^2} \right) 
  - \frac{m_0^4}{8\pi^2} \log
  \left(\frac{\mu^2}{m_0^2} \right)
 + F_1 + F_2
 \;,
\end{equation}
which is shown in Fig.~\ref{fig:nlopressure}. 
At $T=0$, we can calculate $\mathcal{P}_{\mathrm{NLO}} / T^4$ exactly and use
it as a check of the numerical calculations. At $T=0$, clearly $F_1
= 0$, and hence $\mathcal{P}_{\mathrm{NLO}}/T^4 = F_2 / T^4$. 
At $T=0$, it is easy to see that for low $P$, $I(P,m)$ is
dominated by $32\pi^2 \bar \sigma^2 / (P^2 + m^2)$. This gives
\beq
F_2 \approx \frac{1}{2} \sumdiff_P \log(P^2 + m^2) \;. 
\eeq
For $H=0$, the mass $m$ vanishes and so
$\mathcal{P}_{\mathrm{NLO}}/T^4 = -\pi^2 / 90$.
For
$H = (104\;\mathrm{MeV})^3$ we have that 
$\mathcal{P}_{\mathrm{NLO}}/T^4 = 0$,
since $T$ is in that case much smaller than $m$, such that the pressure is
exponentially suppressed.

The pressure for $H=0$ is approaching the $H \neq 0$ pressure at high
temperatures, indicating that 
the effects of the explicit symmetry-breaking terms become
smaller at higher temperatures. This is because $H$ is 
a temperature-independent constant. 

\subsection{Pressure of the $O(4)$ linear sigma model} 
\label{4c}
\begin{figure}[htb]
\begin{center}
\scalebox{0.9}{\includegraphics{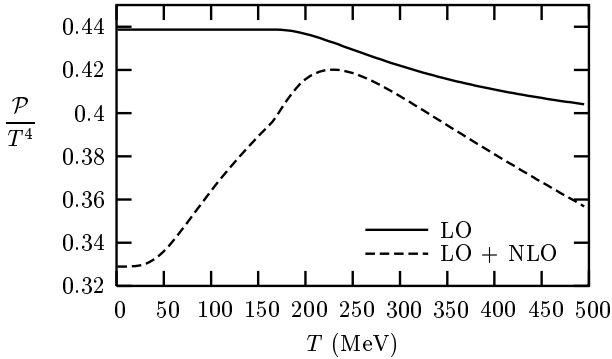}}
\caption{LO and NLO pressure normalized to $T^4$, for $N=4$ as
 a function of temperature, for $H=0$ and $\Lambda =5.0$ GeV.}
\label{fig:presN4h0}
\end{center}
\end{figure}

In order to make contact with two-flavor low-energy QCD, we specialize to 
$N=4$.
In Fig$.$~\ref{fig:presN4h0}, we show the pressure for $N=4$ 
and $H=0$ to next-to-leading order as function of $T$ normalized by
$T^4$.
The LO pressure below $T_c$ equals the pressure of a gas of four
massless noninteracting scalars. 
This follows immediately from Eqs.~(\ref{eq:effpotlo3d}) and~(\ref{Jzero}).
At NLO the sigma field becomes massive. For temperatures much lower 
than $m_{\sigma}$,
the contribution to the pressure from the sigma is Boltzmann 
suppressed and
we have (to good approximation) ${\cal P}=\pi^2 T^4/30$, 
which is the pressure of a gas of three
massless noninteracting scalars. 
From the calculations we conclude that the transition to NLO is of second (or
higher) order since the derivative of the pressure is not diverging. 

\begin{figure}[htb]
\begin{center}
\scalebox{0.9}{\includegraphics{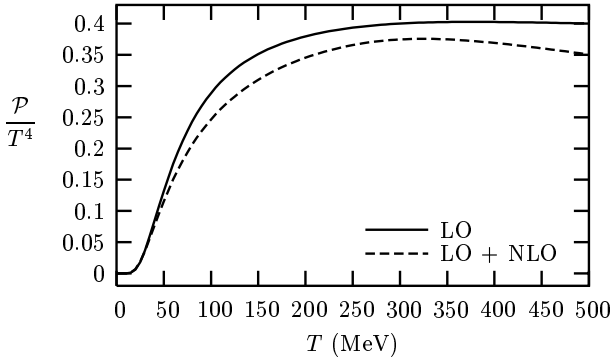}}
\caption{LO and NLO pressure for $N=4$ normalized to $T^4$, as
a function of temperature for $H = \left(104 \;\mathrm{MeV} \right)^3$  
and $\Lambda = 5.0$ GeV.}
\label{fig:presN4h1}
\end{center}
\end{figure}

In Fig$.$~\ref{fig:presN4h1}, we show the pressure for $N=4$ 
and $H = \left(104 \;\mathrm{MeV} \right)^3$
to next-to-leading order as function of $T$ normalized by
$T^4$.

In Figs.\ \ref{fig:presN4h0} and \ref{fig:presN4h1} we have chosen the cutoff 
$\Lambda =5.0$ GeV. Because we wish to make contact with low-energy QCD, a few
comments on this choice are in order. For the low-energy chiral Lagrangian, 
the cutoff is usually taken to be $8\pi f_\pi$ (using our definition of
$f_\pi$), which is around 1.2 GeV. However, for the present purpose this
value would be at the limit of applicability, since the critical temperature 
at which chiral symmetry is (approximately) restored is only about a factor 
of 8 smaller and we have to satisfy the requirement that 
$2 \pi T \ll \Lambda$. In this way one ensures that one sums over
sufficient Matsubara modes \footnote{The amount of Matsubara modes to be
summed over depends on how one implements the cutoff on sum-integrals \cite{Amte}.}. 
Therefore, we have taken the cutoff considerably 
larger to reduce
the sensitivity to the cutoff, but emphasize that only for the region below
$T_c$ can one expect the result to be of relevance for the QCD pressure. 

\subsection{Pressure of the $O(4)$ nonlinear sigma model} 
\label{4d}
In the limit $\lambda_b = \infty$, 
we obtain the Lagrangian for the nonlinear sigma model.
So there are no counterterms for logarithmic divergences. We will only
renormalize $f_{\pi,b}$ as in Eq.~(\ref{eq:fpirenormnlo}) with $\lambda_b
= \infty$. This implies that $F_2$ has a small cutoff dependence. 

\begin{figure}[htb]
\begin{center}
\scalebox{0.9}{\includegraphics{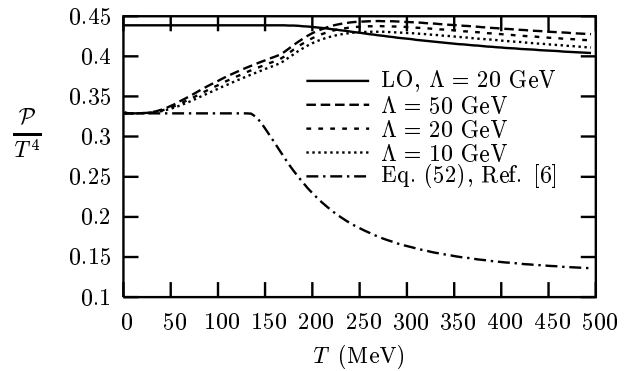}}
\caption{NLO pressure of the nonlinear sigma model for $N=4$ 
normalized to $T^4$, as
a function of temperature for different values of the cutoff
$\Lambda$. For comparison we have included the LO pressure and a curve
corresponding to the NLO 
pressure expression from Ref$.$~\cite{kap}.}
\label{fig:presnon}
\end{center}
\end{figure}

In Fig$.$~\ref{fig:presnon} we show the pressure of the $O(4)$
nonlinear sigma model without explicit symmetry breaking ($H=0$), through
next-to-leading order in $1/N$. We have calculated
the pressure for different values of the cutoff. The LO
result for $\Lambda = 20\;\mathrm{GeV}$ is included. 
For comparison, we also 
show the pressure resulting from the approximations employed  
by Bochkarev and Kapusta, Ref$.$~\cite{kap}. 
A considerable difference
between our results and those of Bochkarev and Kapusta is observed. 

We next discuss the approximations made in Ref$.$~\cite{kap}.
The $T=0$ part of sum-integrals are omitted such that every $\isumint$ 
is replaced by $\isumdiff$. Hence $\Pi(P,m)$ is replaced by $\Pi_T(P,m)$. 
In the
high-energy approximation the latter is approximated by
\begin{equation}
\Pi_T(P,m) \approx \frac{T^2 J_1(\beta m)}{32 \pi^2 A^2},
\end{equation}
where $J_1$ and $A^2$ are 
given by Eqs.~(\ref{J1}) and~(\ref{Akwadraat}), respectively.
The term involving $\bar{\sigma}^2$ in Eq.~(\ref{Ipm}) is omitted
and the pressure reduces to 
\begin{eqnarray}\nonumber
  \mathcal{P} &=& \frac{N m^2}{2}\left( f_{\pi}^2 - \bar \sigma^2 \right) 
        -\frac{N}{2} \sumdiff_P \log(P^2 + m^2 ) \\ 
   & & -\frac{1}{2} \sumdiff_P \log \frac{P^2}{P^4 + 4 m^2 p_0^2} \;.
\end{eqnarray}
Defining the functions 
\begin{equation}
  K_0^{\pm}(\beta m) = \frac{32}{3 T^4} 
\int_0^{\infty}\mathrm{d}p\;\frac{p^4}{\omega_p}\;
  n(\omega_\pm)\;,
\end{equation}
where $\omega_{\pm}=\sqrt{p^2+m^2}\pm m$, the pressure becomes
\begin{eqnarray}\nonumber
  \mathcal{P} &=& \frac{N m^2}{2} \left( f_{\pi}^2 - \bar \sigma^2 \right) 
+ \frac{NT^4 J_0(\beta m)}{64 \pi^2} + \frac{\pi^2}{90} T^4 \\
   & & -\frac{T^4}{64 \pi^2} \left[K_0^+(\beta m) + K_0^-(\beta m) \right] \;.
\label{expnot}
\end{eqnarray}
Expanding Eq.~(\ref{expnot}) in powers of $m/T$
and rescaling with factors of $N$, one
obtains Eq.~(52) of Ref$.$~\cite{kap}.

For completeness, we also give the gap equations in this approximation:
\bqa
16 \pi^2 f_\pi^2 &=& T^2 J_1(\beta m) + 16 \pi^2 \bar \sigma^2 
 \nonumber \\ &&- \frac{T^2}{N} \left[K_1^+(\beta m) + K_1^-(\beta m) \right]
\;,\label{BKgap}  \\
m^2 \bar \sigma &=& 0 \;,
\eqa
where the functions $K_1^{\pm}$ are 
\begin{equation}
  K_1^{\pm}(\beta m) = \pm \frac{8}{T^2} 
\int_0^{\infty}\mathrm{d}p\;\frac{p^2}{\omega_p} \;\frac{\omega_\pm}{m} \; 
n(\omega_\pm)\;.
\end{equation}

There are several problems with the approach in Ref.~\cite{kap}. 
Firstly, it is incorrect
to ignore zero-temperature contributions to the pressure and it also
obscures renormalization issues. Then, as we saw in Sec.~\ref{4b},
the arguments for applying the high-energy approximation are not
valid. Furthermore the term proportional to $\bar{\sigma}$ is not suppressed
by $1/N$ compared to the other terms appearing in $I(P,m)$. 
Fourthly, as the solutions to the gap equation (\ref{BKgap}) 
indicate for $T>T_c$ $m/T$ becomes significantly larger than one, hence the
$m/T$ expansion breaks down. If one were to use Eq.~(\ref{expnot}) 
instead, one finds that the pressure even becomes negative above 
$T \simorder 300$ MeV. 
Another problem is that for $T < T_c$ their
pressure is equal to that of a massless gas. However, this is incorrect
since the sigma meson is massive and included at
NLO. Hence one expects a deviation from the ideal-gas pressure at $T
< T_c$. Finally, at high temperatures we expect that the NLO pressure will become
approximately equal to the LO pressure because chiral symmetry
will be restored. This is not the case for the pressure of Ref.~\cite{kap}.

We briefly comment on the paper by Jain~\cite{jain}.
In that paper the author is calculating the thermodynamic potential
to NLO in the $O(N)$ linear sigma model using a high-temperature
expansion. Since this approximation breaks down at low
temperatures, we will refrain from comparing with our results. 

\section{Choice of parameters}\label{sec:parameters}

In the preceding sections we have shown plots for particular choices of the 
parameters, namely, $\lambda (\mu =
100\;\mathrm{MeV}) = 30$, $f_\pi = 47\;\mathrm{MeV}$ and if 
there is explicit
symmetry breaking, we take
 $H = \left(104\;\mathrm{MeV} \right)^3$. In this section
we will motivate these choices. For simplicity we partly use leading-order
calculations for fixing the parameters. 

We start with choosing the values for $f_\pi$ and $m_\pi$ to be roughly equal
to their measured values: 
$f_\pi = 47\;\mathrm{MeV}$ (note that our $f_\pi$ 
differs from the more conventional definition by a factor of $1/2$) 
and $m_{\pi}= 138$ MeV (the
average of the measured masses of the $\pi^0, \pi^+$ and $\pi^-$). 
We will use this for choosing our parameter $H$ as follows. 
Given a choice of $\lambda$ at a given scale $\mu$ we solve 
the LO renormalized 
gap equations (\ref{eq:logapeqm}) and (\ref{eq:logapeqsigma}), for $\bar 
\sigma$ and $m^2$, such that $m^2 = m_\pi^2$ (which is the correct
identification at LO). For our choice of $\lambda(\mu =
100\;\mathrm{MeV}) = 30$, this results in 
$H = \left(104 \;\mathrm{MeV} \right)^3$.

The choice of $\lambda$ is motivated by
considerations on the maximal value of the cutoff and the sigma mass.
To obtain this mass, one has to find the 
poles of the propagators in Minkowski space. 
The physical mass $m_{\mathrm{ph}}$ is often defined by the solution to 
the equation
\beq
  -m_{\mathrm{ph}}^2 + m^2 + \mathrm{Re}{\Sigma(p_0 = i m_\mathrm{ph}
   + \epsilon,\,p = 0,\, m)} = 0\;,
\eeq
where $\Sigma$ is the self-energy.
Using Eq.\ (\ref{inversesigmaprop}) and choosing $\mu =
m_\sigma$, we find that at $T=0$ and for $H=0$
\begin{equation} 
  m^2_\sigma = \frac{32
\pi^2 f_\pi^2}{1 + \frac{32 \pi^2}{\lambda(m_\sigma)} + \frac{\pi^2}{1
+ 32\pi^2 / \lambda(m_\sigma)}} \;.  
 \label{eq:sigmamass}
\end{equation} 
Equation (\ref{eq:sigmamass}) implies that
$m_\sigma \leq \sqrt{16 \pi} f_\pi \approx 333$ MeV, which is lower than 
600 MeV. For $H \neq 0$ a similar bound
applies. In that case we find that the maximal value of the sigma mass
can be found by solving the following equation for $m_\sigma$
\beq
  m_\sigma^2 = \left[2 + \sqrt{1+ A^2(m_\sigma^2 /
  m_\pi^2)}\right] m_\pi^2 \;,
\eeq
where
\bqa
  A(x) &=& \left(\frac{16\pi f_\pi^2}{m_\pi^2} - \frac{1}{\pi} \right)
  \frac{1}{\sqrt{1- 4 / x}} \nonumber \\
  & & - \frac{1}{\pi} \log \left( \frac{1 - \sqrt{1 - 4/x}}{1+ \sqrt{1 - 4 /x}}
\right)
  \;.
\eqa
We find by solving this equation that the maximal value of $m_\sigma$
is equal to 433 MeV. This is also smaller than the average measured
value of 600 MeV. The reason that we find an unphysical bound for the
sigma mass could be that we consider $N_f=2$ and may miss out on
essential three-flavor physics.

The sigma mass turns out to be maximal if $\lambda(\mu =
100\;\mathrm{MeV}) = 80$. The problem with this choice of $\lambda$ is
that in that case the maximal value of the cutoff is 720 MeV. This is
very low and
allows us only to do calculations up to around $T =
50\;\mathrm{MeV}$. Therefore we choose a lower value: 
$\lambda(\mu = 100\;\mathrm{MeV}) = 30$. Using that parameter choice $\Lambda_{\mathrm{max}} =
19\;\mathrm{GeV}$ and the sigma mass is equal to 256 MeV and 350 MeV
in the case of $H = 0$ and $H = (104\;\mathrm{MeV})^3$ respectively.

\section{Summary and Conclusions}
In this paper, we have considered the thermodynamics of the $O(N)$
linear and nonlinear sigma models to NLO in the $1/N$ expansion.

At NLO we have shown that one can renormalize the effective potential
in a temperature-independent manner only at the minimum of the
effective potential.  This is another example of the ambiguity
in the definition of off-shell Green's functions. A perhaps more
familiar example comes from the calculation of the zero-temperature
effective potential in gauge theories. In this case, the effective
potential depends on the gauge-fixing condition except at the
minimum~\cite{jackiw,nkn,kobes}.
By renormalizing the NLO effective potential in the minimum we found
the beta function for $\lambda$ to NLO. This beta function is
consistent with the perturbative calculation.

We calculated numerically the pressure for the linear and nonlinear
sigma model to NLO as a function of
temperature. Our calculations show that for the calculation of the
pressure $1/N$ is a good expansion, even if $N=4$. With a relatively realistic
choice of the parameters we made a prediction for the pressure of QCD
at low temperatures.
Our results for the pressure disagree significantly with the
calculations of those in Ref~\cite{kap}.  This is due to the fact that
we are not neglecting zero-temperature contributions and that we treat
the NLO contribution without resorting to any high-energy
approximation.

We also found that in the linear sigma model the sigma mass has an
upper bound. This bound depends only on the parameters
$f_\pi$ and $m_\pi$. For a
realistic choice of these parameters, this implies that the sigma mass
is smaller than 433 MeV. This does not necessarily have consequences for
the real sigma meson, since we did not take into account the full
three-flavor physics. 

Having solved the $O(N)$ model for the thermodynamics, it is natural
to apply it to other quantities such as spectral functions. The
methods developed here should also be useful for more complicated
models incorporating additional features of low-energy QCD, 
e.g. $U(3)_A\times U(3)_V$.

\vspace{-1.0cm}

\section*{Acknowledgments} 
The research of D.B.~has been made possible by financial support from the 
Royal Netherlands Academy of Arts and Sciences.

\end{document}